\documentclass[12pt,preprint]{aastex}

\newcommand{\msol}{\,$M_{\odot}$}

\begin{document}

\title{Four Brown Dwarfs in the Taurus Star-Forming Region}

\author{E. L. Mart\'{\i}n\altaffilmark{1}, 
C. Dougados\altaffilmark{2}, E. Magnier\altaffilmark{1,2}, F.  M\'enard\altaffilmark{2}, 
A. Magazz\`u\altaffilmark{3,4}, J.-C. Cuillandre\altaffilmark{2}, X. Delfosse\altaffilmark{5}} 

\altaffiltext{1}{ Institute of Astronomy. University of Hawaii, 
2680 Woodlawn Drive. Honolulu, HI 96822, USA}
\altaffiltext{2}{Canada-France-Hawaii Telescope Corporation, P.O. Box 1597, 
Kamuela, HI 96743, USA}
\altaffiltext{3}{Centro Galileo Galilei, Apartado 565, E-38700, Santa Cruz de La Palma, Spain}
\altaffiltext{4}{Osservatorio Astrofisico di Catania, Via S. Sofia 78, I-95123 Catania, 
Italy}
\altaffiltext{5}{Laboratoire d'Astrophysique de Grenoble, BP 53, 38041 Grenoble, France}




\begin{abstract}

We have identified four brown dwarfs in the 
Taurus star-forming region. They were first selected from $R$ and $I$ 
CCD photometry of 2.29 square degrees obtained at the 
Canada-France-Hawaii Telescope. Subsequently, they were recovered 
in the 2MASS second incremental data release point source catalog. 
Low-resolution optical spectra obtained at the William Herschel telescope 
allow us to derive spectral types in the range M7--M9. 
One of the brown dwarfs has very strong  H$\alpha$ 
emission (EW=-340~\AA\ ). It also displays Br$\gamma$ emission in an 
infrared spectrum obtained with IRCS on the Subaru telescope, suggesting that  
it is accreting matter from a disk.  
The \ion{K}{1} resonance doublet and the \ion{Na}{1} subordinate doublet at 818.3 and 819.5~nm in 
these Taurus objects are weaker than 
in field dwarfs of similar spectral type, consistent with 
low surface gravities as expected for young brown dwarfs. 
Two of the objects are cooler and fainter than GG~Tau~Bb, the  lowest 
mass known member of the Taurus association. 
We estimate masses of only 0.03~M$_\odot$ for them. 
The spatial distribution of brown dwarfs in Taurus hints to a possible anticorrelation 
between the density of stars and the density of brown dwarfs. 
 
\end{abstract}


\section{Introduction}

The Taurus star-formation region (SFR) 
has played a central role in our understanding of low-mass 
star formation. The dark molecular clouds in this region of the sky are 
conspicuously large. Early work classified them as 
``nebulae without definite relation to certain stars`` 
\citep{cederblad46}. 
The class of very young low-mass variable stars known as T Tauri stars (TTS) is named after 
the prototype star T Tauri \citep{joy42}, which is a member of the Taurus SFR. 
The importance of the Taurus SFR is that it contains many nearby TTS (distance $\sim$ 140~pc) 
in loose aggregations. The ages of 
Taurus members range from protostars to evolved post T Tauri stars, although the bulk 
of them appears to have emerged from the molecular clouds in the last 4~Myr 
\citep{palla00}. 

Recent surveys in very young clusters have identified a population of substellar 
objects ($M\,<\,0.075$\msol; \cite{kumar63}) that appears to be of the same order 
of magnitude in number as that of stars \citep{bejar01,bouvier98}. 
Several brown dwarfs have   
been identified in SFRs such as the Chamaeleon~I SFR \citep{comeron00} and the 
Trapezium \citep{lucas01}. 
The Initial Mass Function (IMF) does not appear to change much from one region to another. 
Taurus is interesting because it represents a loose mode of star formation 
(density 1--10 pc$^{-3}$), in contrast to that of clusters such as the 
Trapezium (10$^4$ pc$^{-3}$).  
Recent searches in Taurus by \citet{briceno98} and \citet{luhman00} 
have failed to reveal any candidate members with spectral types 
later than M6.5V, estimated to be below the substellar limit at the Taurus population
age \citep{martin99}. \citet{luhman00} have proposed that the low-mass IMF 
in Taurus could be truncated around the substellar limit. 
It is indeed important to study whether the IMF could be sensitive to the initial conditions 
of star formation. 

In this paper we present the first results of a new effort to search for brown dwarfs 
in Taurus. We have obtained deeper images over a 5 times larger area than previous work. 
Spectroscopic follow-up of 30 very low-mass (VLM) candidate members in the Taurus SFR has 
been carried out. Four candidates are confirmed as VLM objects with spectral types M7 or later, and  
the probability of membership in the Taurus SFR is very high. 
They are sufficiently cool to be considered as 
brown dwarfs. One of them presents strong evidence for disk accretion.

\section{Observations and results}

Direct imaging observations of 11 fields in the Taurus SFR were
obtained with the 3.6-m Canada-France-Hawaii Telescope (CFHT) CFH12k
camera \citep{cuillandre00} 
between 1999 December and 2000 December. Broad band $R_{\rm C}$, $I_{\rm C}$ and $z$
as well as narrow band H$\alpha$ filters were used. The field
of view of CFH12k is 0.327 square degrees. Thus our survey
covers a total area of 3.59 square degrees. A complete description of
the survey and associated data reduction will be presented in Dougados
et al. (2001, in preparation). The selection of Taurus low-mass candidates
presented here is based on a subset of the total CFH12k survey, covering
2.29 square degrees (7 fields) and including $R_{\rm C}$ and $I_{\rm C}$ 
photometry alone. 

We selected candidate low-mass Taurus members by requiring that they
lie 2 magnitudes above the observational ZAMS in the $I$ versus $R$-$I$ 
color-magnitude diagram (20 sources). An additional  10 sources lying 
closer to the ZAMS with suspected strong H$\alpha$ emission from their photometry in the 
H$\alpha$ filter were included in the spectroscopic follow-up sample. 
None of the H$\alpha$ selected sources turned out 
to be interesting. 
Figure~1 shows the color-magnitude diagram. 

Spectroscopic observations were carried out using ISIS at the 4.2-m William Herschel 
Telescope (WHT) in La Palma on 2000 September 28-29. 
We observed 30 VLM candidates with $I<$18. V410~Tau~Anon~13 (M6.5; \cite{briceno98}) 
was also observed with the same setup for comparison purposes. 
The R158R grating on ISIS's red arm 
gave a wavelength coverage from 640.9~nm to 936.5~nm. The spectral resolution was 2.5 pixel 
(7.2~\AA\ ). The data were reduced using standard routines  
for bias-subtraction and flat-field correction within the IRAF\footnote{IRAF is distributed 
by National Optical Astronomy Observatories, which is operated by the Association of 
Universities for Research in Astronomy, Inc., under contract with the National 
Science Foundation.} enviroment. Wavelength calibration was made using the spectrum 
of a NeAr lamp. 
Instrumental response was calibrated out using spectra of the flux standard Feige 24. 
 
This paper focuses on the four new VLM candidates for which our WHT spectra give spectral types 
M7 or later (Figure~2). These objects represent an extension of the previously known 
Taurus members to lower masses, beyond the substellar limit. 
Some of the remaining 26 candidates have H$_\alpha$ emission and spectral types earlier than M7. 
They could be Taurus SFR members. We defer presentation of these stars to 
another paper (Dougados et al. 2001, in preparation). 
Spectral types were obtained by measuring the strength of VO absorption using 
indexes defined by \citet{martin99} calibrated with spectra of 
field ultracool dwarfs. We obtain a spectral type for V410~Tau~Anon~13 of M6, consistent within 
the uncertainties (half a subclass) with the M6.5 estimate of \citet{briceno98}. 

Near-infrared photometry for the program objects was extracted 
from the 2MASS second incremental data release point source catalog. 
Line-of-sight extinction was measured using the $I-J$ colors of field M dwarfs and 
the interstellar  extinction law of \citet{rieke85}.   
The astrometric, photometric and spectroscopic information 
for these sources is summarized in Tables 1 and 2. 

We obtained near-infrared spectra for two of these objects using the Infrared Camera and Spectrograph 
(IRCS; \cite{kobayashi00}) 
at the Cassegrain focus of the 8-m Subaru telescope on 2001 January 9. The $K$-band grism gave 
a wavelength coverage from 1.93 to 2.48~$\mu$m, a dispersion of 6.1~A~pix$^{-1}$ and a resolution 
of 350 at 2.2~$\mu$m. The data were reduced using standard IRAF routines  
for bias-subtraction, flat-field correction and sky-subtraction. Telluric bands were cancelled 
using the spectrum of the G3 star SAO~167029. 
The photospheric Br$\gamma$ line in  SAO~167029 was 
removed by interpolation between adjacent continuum points. The spectra of the targets were 
multiplied by a blackbody with T$_{\rm eff}$=5785~K, adequate for the G3 spectral type of 
the standard star \citep{tokunaga00}. 
The final IRCS spectra are shown in Figure~3.

\section{Properties of the new Taurus VLM members}

Brown dwarfs in the Taurus SFR should be in a very early phase of gravitational contraction. 
Theoretical models indicate that their gravities are between 10 to 100 times lower than 
for VLM stars at the 
bottom of the main sequence \citep{chabrier00}.  Collisionally dominated 
photospheric lines are thus very good tracers of low gravity atmospheres and can be used 
as a strong criterion to distinguish young BDs from old VLM stars. 
\citet{luhman98} noted that \ion{Na}{1} and \ion{K}{1} lines were much weaker in the M6 Taurus star 
V410~X-ray~3 than among field M6 dwarfs. We find that our four new Taurus VLM objects 
also have weaker \ion{Na}{1} and \ion{K}{1} lines than field dwarfs of the same spectral type 
(Figures 2 and 3). 

A spectral type of M6.5 is considered the boundary 
between stars and brown dwarfs for  the Pleiades cluster age, i.e. 120~Myr. 
For younger ages, the boundary 
should stay at M6.5 or move to slightly earlier spectral type 
\citep{martin99}. Thus, we conclude that the four objects presented 
in this letter are very likely substellar members in the Taurus SFR because 
they have spectral types M7 or later and low-gravity indications in their spectra. 
Prior to this 
work, the coolest known Taurus member was GG~Tau~Bb, the faintest member of the GG~Tau 
quadruple system.  \citet{white99} estimated 
an age of 1.5~Myr and a mass of 0.04~\msol for it. 
GG~Tau~Bb has spectral type M7, $K$=12.01 and A$_{V}\sim$ 0. Two of our objects, namely 
CFHT-BD-Tau~2 and 3, are intrinsically 
fainter than GG~Tau~Bb. Using the models of \citet{chabrier00}, 
which provide the best isochrone fitting to the GG~Tau quadruple system,  
we obtain ages of about 1~Myr, and masses of about 0.03~\msol for both 
CFHT-BD-Tau~2 and 3. CFHT-BD-Tau~1 and 4 appear to be more luminous than the 
1~Myr isochrone. Dougados et al. (2001, in preparation) will present a detailed 
discussion of the position of these objects in the H-R diagram.

Our IRCS data allows us to search for Br$\gamma$ in emission, which has been detected in classical TTSs  
\citep{muzerolle98}. These authors found a correlation between the strength  
of Br$\gamma$ emission and the accretion rate estimated from $U$-band excess and blue excess 
spectra of TTSs. We find Br$\gamma$ emission in CFHT-BD-Tau~4 (Figure~3), which also displays the 
strongest H$\alpha$ emission in our sample (Table~2). The analogy with the TTSs suggest that 
CFHT-BD-Tau~4 may be accreting matter from a circumstellar disk. 
 
V410~Tau~Anon~13 is a VLM Taurus member with H$\alpha$ emission 
(EW(H$\alpha$)=-41.3~\AA ; \citet{briceno98}). We measured a similar 
strength of H$\alpha$ emission (EW(H$\alpha$)=-35~\AA ) in our WHT spectrum. 
Our Br$\gamma$ detection for V410~Tau~Anon~13 confirms that this object may be accreting 
matter as previously proposed by \citet{muzerolle00} from a high-resolution 
study of the H$\alpha$ line profile. The continuum of the IRCS spectra of CFHT-BD-Tau~4 and 
V410~Tau~Anon~13 are significantly redder and the CO bands are shallower 
than in the comparison spectrum of LHS 3003, contrarily to what 
one would expect from the sensitivity of these bands to the 
surface gravity. This could be due to emission from the circumstellar disk, which could 
add a red continuum and could veil the CO features. 
 
CFHT-BD-Tau~4 is located close to the Tau~III group identified by 
\citet{gomez93}. CFHT-BD-Tau~1, 2 and 3 could be associated with group V, 
for which only 
half a dozen members were known so far. Group V has much fewer known T Tauri members than 
groups II and III. We note that three out of four of our new  
BDs are located in a sparsely populated group, and none in one of the richer groups I, II and III, 
which were included in the photometric survey. 
The spatial distribution of 
BDs in Taurus may give an important clue to the dominant formation mechanism of these 
objects. Our results are intriguing because they suggest that the density of BDs in the Taurus SFR 
may be anticorrelated with the density of TTSs. \citet{martin96} found 
an isolated group of VLM young stars at high galactic latitude and suggested that VLM 
stars and brown dwarfs may form in small groups without any low-mass star. 
A mass segregation in Taurus would 
explain the lack of brown dwarfs in the searches carried out by \citet{briceno98} 
and \citet{luhman00}, which were limited to the dense groups of TTSs. 
But why is there a lack of BDs in dense groups of TTSs in the Taurus SFR? 
A possible answer may be that in rich groups BDs get ejected at a very early 
stage \citep{reipurth01} and disperse 
away from the molecular clouds in a short time. The binary frequency of TTSs in the Taurus SFR is higher 
than in other SFRs  \citep{leinert93}, such as the Trapezium  \citep{prosser94}, 
implying that the ejection mechanism may be more efficient in Taurus. 
Fully resolving this issue will however require
studying a larger sample of Taurus BDs, which the completion of the CFH12k
survey should soon provide.

\acknowledgments

The William Herschel Telescope is operated on the island of La Palma 
by the Isaac Newton Group in the Spanish Observatorio del Roque de los 
Muchachos of the Instituto de Astrof\'\i sica de Canarias. 
We thank G. Herbig for comments on the early work on star formation in Taurus. 
Partial financial support was provided by NASA through a grant 
from the Space Telescope Science Institute, which is operated by the Association of 
Universities for Research in Astronomy, Inc., under NASA contract NAS5-26555.

\clearpage

\figcaption[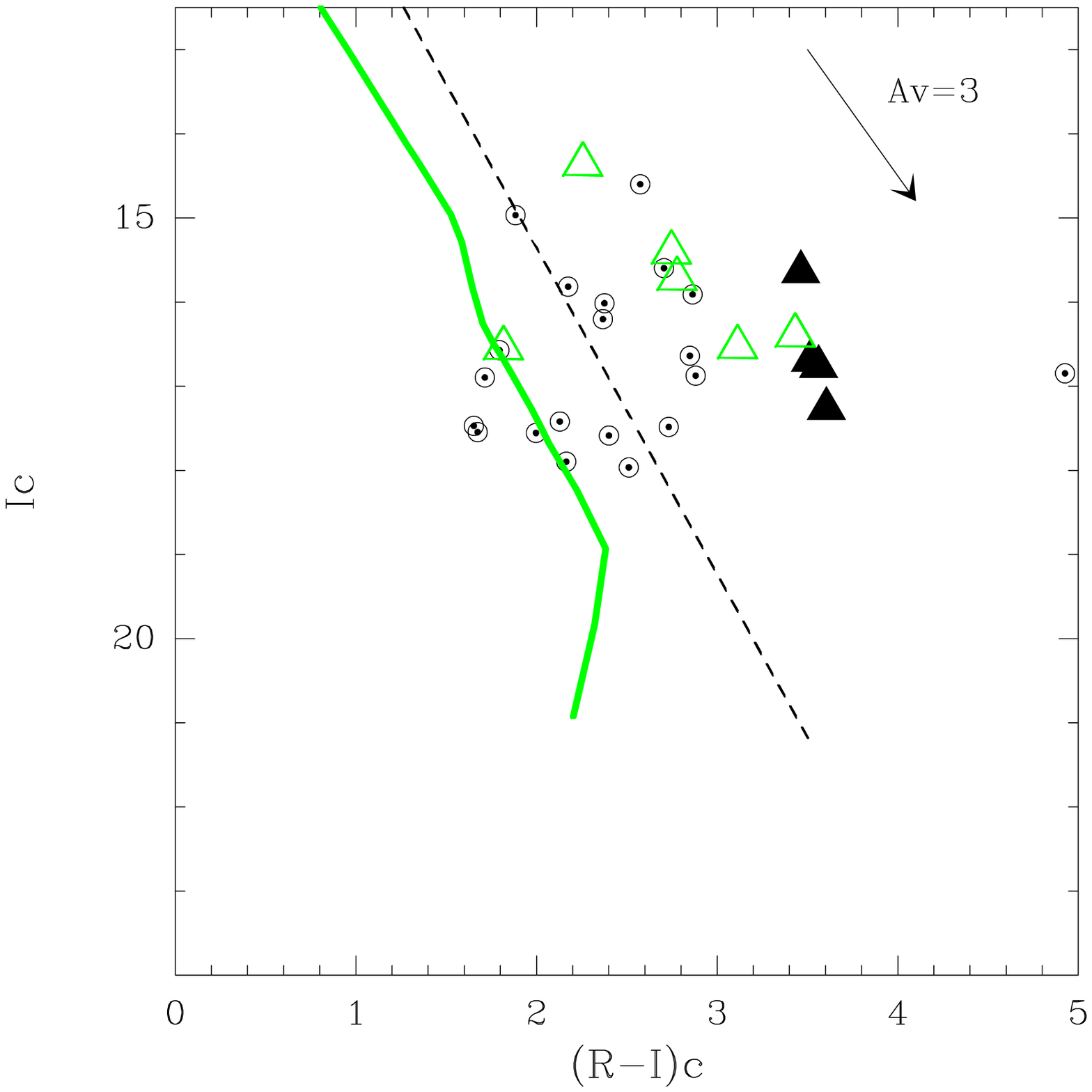]{\label{fig1} $I_{\rm C}$ vs. $(R-I)_{\rm C}$ color--magnitude 
diagram for CFH12K data. Small dots surrounded by circles are 
background stars. Filled triangles are 
spectroscopically confirmed new very low-mass Taurus members. Empty triangles are previously 
known Taurus members recovered in the CFH12K data. The solid line is the locus of 
field M dwarfs taken to the Taurus distance \citep{weis84,leggett92}. 
The dashed line represents an observational boundary between 
Taurus members and background stars. There is only one member bluer than this line.} 

\figcaption[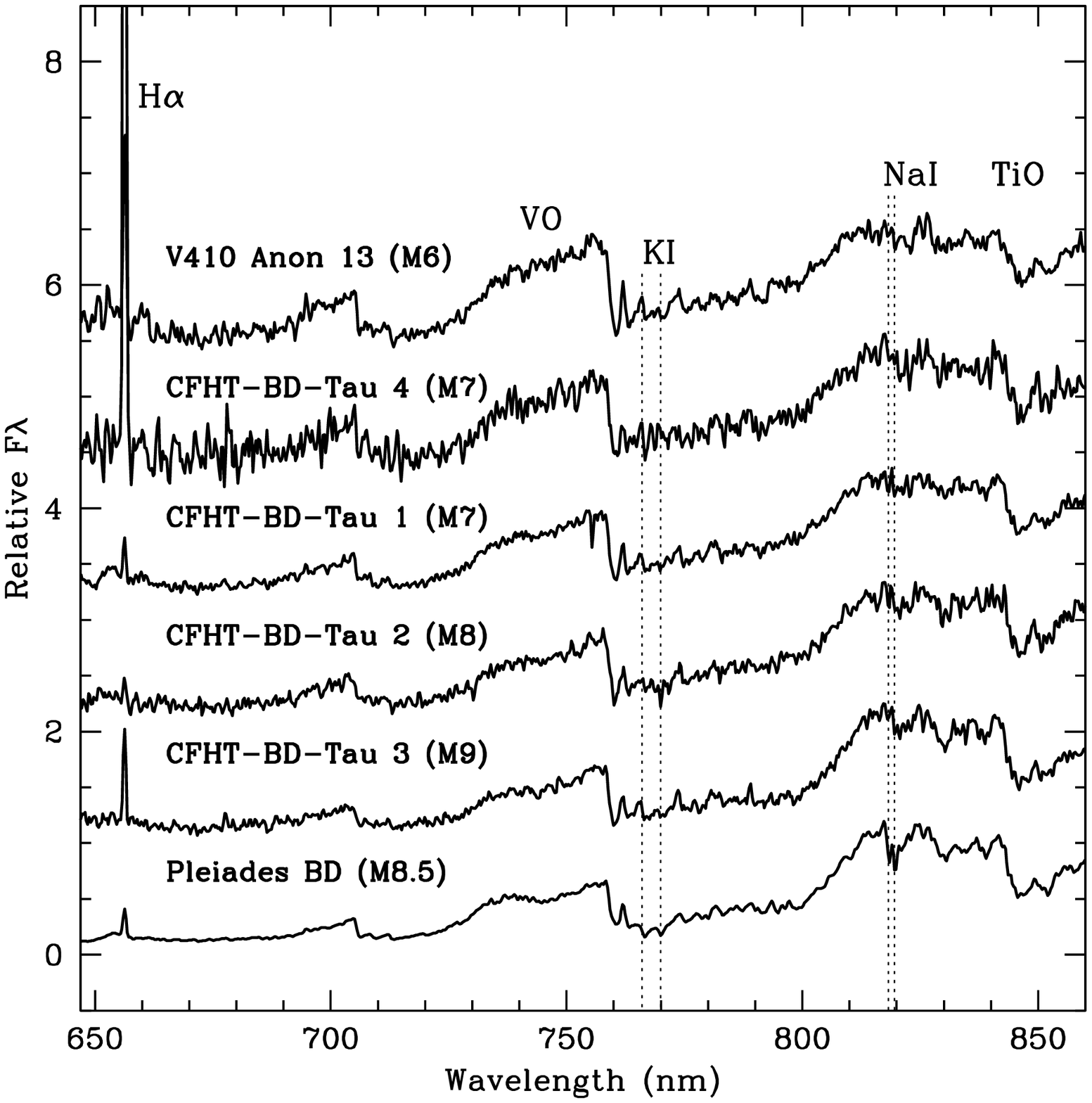]{\label{fig2} Final WHT spectra for CFHT Taurus BD members. 
The top spectrum is a known M6 Taurus member \citep{briceno98}. 
The bottom spectrum is an average of the Pleiades 
BDs Teide~1 (M8) and Roque~4 (M9). The spectra of Taurus objects 
have been dereddened using the A$_V$ values given in Table~1. 
Note that the \ion{K}{1} resonance doublet and the \ion{Na}{1} subordinate doublet at 818.3 and 819.5~nm 
are weaker in the Taurus BDs than in the Pleiades brown dwarf.} 

\figcaption[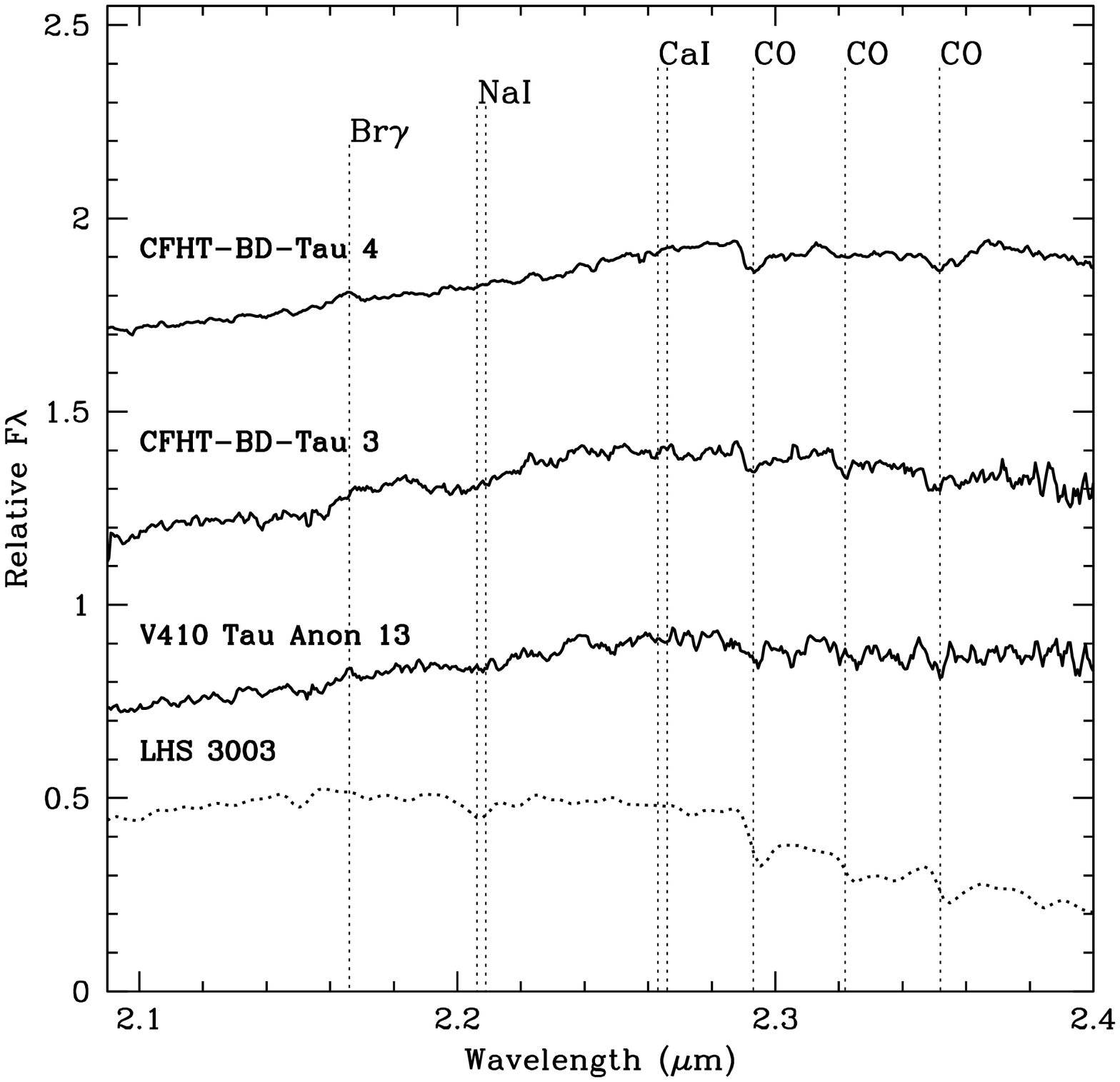]{\label{fig3} 
Final IRCS K band spectra of two new Taurus brown dwarfs and V410~Tau~Anon~13. 
A boxcar smoothing of 3 pixels has been applied to the IRCS spectra. For comparison we also 
plot the spectrum of the M7 dwarf LHS~3003 (dotted line) taken from  \citet{leggett01}. 
Note that the \ion{Na}{1} line is weaker in the Taurus objects than in the field M7 dwarf.}

\clearpage
\epsscale{1.1}
\plotone{fig1.eps}
\clearpage
\epsscale{1.1}
\plotone{fig2.eps}
\clearpage
\epsscale{1.1}
\plotone{fig3.eps}
\clearpage

\clearpage
\begin{deluxetable}{lccccccc}
\scriptsize
\tablecaption{\label{tab1} Astrometric and photometric data for program objects}
\tablewidth{0pt}
\tablehead{\colhead{Name (IAU)} & \colhead{Name (short)} & 
\colhead{R$_{\rm C}$} & \colhead{I$_{\rm C}$} & \colhead{J} & \colhead{H} &  \colhead{K} & \colhead{A$_V$}  }
\startdata
CFHT-BD-Tau J043415.2+225031 & CFHT-BD-Tau~1 & 20.87 & 17.26 & 13.74 & 12.52 & 11.87 & 3.1 \nl
CFHT-BD-Tau J043610.4+225956 & CFHT-BD-Tau~2 & 20.21 & 16.69 & 13.76 & 12.76 & 12.19 & 0.0 \nl
CFHT-BD-Tau J043638.9+225812 & CFHT-BD-Tau~3 & 20.33 & 16.77 & 13.70 & 12.84 & 12.34 & 0.0 \nl
CFHT-BD-Tau J043947.3+260139 & CFHT-BD-Tau~4 & 19.10 & 15.64 & 12.16 & 11.01 & 10.33 & 3.0 \nl
                             & V410~Tau~Anon~13  & 19.83 & 16.40 & 12.90 & 11.66 & 10.94 & 3.6 \nl
\enddata
\tablecomments{The IAU name contains the equatorial coordinates, equinox J2000. 
The accuracy of the coordinates is better than 1 arcsec. V410~Tau~Anon~13 is included 
for comparison.}
\end{deluxetable}

\clearpage
\begin{deluxetable}{lcccccc}
\scriptsize
\tablecaption{\label{tab1} Spectroscopic data for program objects}
\tablewidth{0pt}
\tablehead{\colhead{Name (short)} &  
\colhead{EW(H$\alpha$)} & \colhead{EW(\ion{Na}{1})} & \colhead{EW(Br$\gamma$)} & 
\colhead{TiO-index} & \colhead{VO-index} & \colhead{SpT} }
\startdata
CFHT-BD-Tau~1     & -19$\pm$4   & $<$2.0 &         & 3.57 & 2.66 & M7  \nl
CFHT-BD-Tau~2     & -13$\pm$4   & $<$1.4 &         & 3.80 & 2.70 & M8  \nl
CFHT-BD-Tau~3     & -55$\pm$4   & $<$2.2 & $>$-1.2 & 4.32 & 2.94 & M9  \nl
CFHT-BD-Tau~4     & -340$\pm$20 & $<$1.1 & -3.0    & 3.20 & 2.65 & M7  \nl
V410~Tau~Anon~13  & -35$pm$2    & $<$1.5 & -1.6    & 3.31 & 2.52 & M6  \nl
\enddata
\tablecomments{The equivalent widths (EW) are given in \AA . Spectral types 
are accurate to half a subclass. 
The \ion{Na}{1} refers to the combined subordinate doublet at 818.3 and 819.5~nm. 
TiO and VO are the sum of the TiO1 and TiO2 indexes, 
and the VO1 and VO2 indexes, respectively, defined by \citet{martin99}.}
\end{deluxetable}

\end{document}